\documentstyle{mn}
\input epsf

\begin{document}

%\journal{astro-ph/000}
\title[Redshift space bias and $\beta$ from the halo model]{Redshift space bias
and $\beta$ from the halo model}
\author[Uro\v s  Seljak]{Uro\v s Seljak\\
Department of Physics, Jadwin Hall\\ Princeton University,
        Princeton, NJ 08544 
}
\date{August 2000}
\pubyear{2000}
\maketitle

\begin{abstract}
We analyze scale dependence of redshift space bias $b$
and $\beta \equiv \Omega_m^{0.6}/b$ in the context of the 
halo model. 
We show that linear bias is a good approximation only on large scales,
for $k<0.1h$Mpc$^{-1}$. 
On intermediate scales
the virial motions of galaxies 
cause a suppression of the power spectrum relative to the linear one 
and the suppression differs from the same effect in dark matter.
This can potentially mimic the effect 
of massive neutrinos and the degeneracy can only be broken if power spectrum 
is measured for $k \ll 0.1h$Mpc$^{-1}$. 
Different methods to determine $\beta$
converge for $k<0.1h$Mpc$^{-1}$, but give drastically different results 
on smaller scales, which explains some of the trends observed in the real 
data. We also asses the level of stochasticity by calculating 
the cross-correlation coefficient between the reconstructed
velocity field divergence and the galaxies and show that 
the two fields decorrelate
for $k>0.1h$Mpc$^{-1}$. 
Most problematic are galaxies predominantly found in groups
and clusters, such as bright, red or elliptical galaxies, where we find poor
convergence to a constant bias or $\beta$ even on large scales.

\end{abstract}

\section{Introduction}

Determination of the power spectrum of mass fluctuations is one of the 
main goals of existing and upcoming galaxy surveys. Current state of the 
art is PSCz \cite{Sau}, which has a near spherical geometry and consists of 
about 15000 measured galaxy redshifts. Upcoming surveys, such 
as 2 degree Field survey (2dF)\footnote{http://www.mso.anu.edu.au/2dFGRS} and 
Sloan Digital Sky Survey (SDSS) 
\footnote{http://www.astro.princeton.edu/BBOOK}, 
will measure redshifts of up to a million galaxies. 
3-dimensional mass power spectrum is sensitive to a number of 
cosmological parameters, such as the matter and baryon density, shape 
and amplitude of initial fluctuations and the Hubble constant. 
This sensitivity is further improved if 
additional information from cosmic microwave background (CMB) anisotropies
is included  \cite{Eise}. Mass power spectrum is particulary important 
for determination of neutrino mass. Massive neutrinos have only a
minor impact on the CMB, but they strongly suppress the level of 
mass fluctuations on small scales because of the high neutrino momentum
before they become nonrelativistic.
In principle the sensitivity of upcoming surveys is such that it 
will be possible to test neutrino masses below 0.1-1eV \cite{het98}, 
close to those
suggested by recent 
Super-Kamiokande neutrino results \cite{kamio98}. A
possible 
concern is that the effect of massive neutrinos becomes important on 
small scales, where the assumption of galaxies tracing dark matter may 
not hold. One of the purposes of this paper is to investigate how serious this
problem is and, more generally, what is the relation between the 
observed redshift space galaxy power spectrum and the underlying linear 
dark matter spectrum.  

The relation between the galaxies and the dark matter clustering has recently 
been analyzed in the context of the halo model \cite{Sel,PeaSmi,Sco}.
In this model all the mass in the universe is divided up into halos
of different mass. These halos cluster
according to the linear theory, up to an overall amplitude which 
depends on the halo mass (halo biasing). To these correlations 
important on large scales one adds correlations on small scales,
which arise from within the same halos. For the latter one needs
to specify the radial halo profile, which can also be a function 
of halo mass. This approach has been succesful in reproducing the 
nonlinear dark matter power spectrum and its transition to the 
linear regime \cite{Sel,Sco,MaFry}. 

Galaxies differ from the dark matter
in their galaxy multiplicity function, which parametrizes the 
number of galaxies inside the 
halo as a function of halo mass. For a magnitude limited sample 
this function is zero for low mass
halos which cannot host bright $L_*$ galaxies, which already implies
that galaxies cannot trace dark matter exactly. Above the threshold the number 
of galaxies increases with the halo mass, but 
need not grow linearly, as suggested by the gas cooling arguments where 
gas in more massive and thus hotter halos takes longer to cool and form 
stars.
In addition to the multiplicity function there is another effect that 
changes the galaxy clustering properties: one galaxy 
is expected to form at the halo center, which enhances the correlations
on small scales. These features naturally explain many of the 
observational properties of galaxy clustering in real space, such as the 
power law growth on small scales and the delayed onset of 
nonlinear clustering in the translinear regime \cite{Sel,Sco,PeaSmi}. 

Previous work in the halo model 
has focused on the clustering in real space, while most 
of existing and upcoming surveys operate in redshift space.
Redshift space distortions enhance the correlations on large 
linear scales and suppress them on small scales. This changes 
the clustering pattern in a nontrivial way and 
it is therefore important to include these effects when 
studying the relation between the galaxy and the dark matter 
power spectrum. As an added bonus, redshift 
distortions also allow one to determine $\beta=\Omega_m^{0.6}/b$,
where $\Omega_m$ is the mass density and $b$ the bias parameter
of the galaxies. There are several existing methods to determine
this parameter in the literature (see \cite{StraWill} for a review).
In this paper we address whether different methods 
to determine $\beta$ converge and what is the survey size needed for this. 

\section{The halo model}

The halo model uses the Press-Schechter \cite{PreSch} picture
for dark matter, which 
assumes all the matter is in a
form of isolated halos with a well defined mass $M$ and halo profile
$\rho(r,M)$, which can be modelled as 
\begin{equation}
\rho(r)={\rho_s \over (r/r_s)^{-\alpha}(1+r/r_s)^{3+\alpha}},
\label{rho}
\end{equation}
where N-body simulations give $1<\alpha<1.5$
\cite{NFW,Moore}.
For the power spectrum analysis adopted here
it is convenient to introduce the Fourier 
transform of the halo profile, normalized to unity on large 
scales,
\begin{equation}
y(k)={1 \over M} \int \rho(r) {\sin (kr) \over kr} d^3r.
\end{equation}
The mass is determined by the total mass within the virial
radius $r_v$,
defined to be the radius where the mean density within it is 
$\delta_{\rm vir}=200$ times
the mean density of the universe. The concentration parameter $c=r_v/r_s$ 
in general depends on the halo mass. In this paper we will use 
$\alpha=1.5$ and $c(M)=6(M/M_*)^{-0.15}$, 
where $M_*$ is the nonlinear mass scale defined below. 
This choice fits well the 
results of N-body simulations \cite{Moore} and has been shown to 
give good agreement with real space power spectra from 
N-body simulations \cite{Sel}, but 
we note that other fits with $\alpha=1$ and a different choice of 
$c(M)$ can give equally good agreement with these. 

In the halo model the power spectrum consists of two terms.
The first is that due to a system of correlated halos,
with inter-halo correlations assumed to be a biased sampling of
$P_{\rm lin}(k)$.
Since the real space convolution is simply a Fourier space multiplication this
contribution is
\begin{equation}
P^{hh}(k) = P_{\rm lin}(k) \left[
  \int f(\nu) d\nu\ b(\nu) y(k; M) \right]^2
\label{eqn:twohalo}
\end{equation}
where $b(\nu)$ is the (linear) bias of a halo of mass $M(\nu)$ and $f(\nu)$
is the multiplicity function.  The peak height $\nu$ is related to the mass
of the halo through
\begin{equation}
  \nu \equiv \left( {\delta_c\over \sigma(M)} \right)^2
\end{equation}
where $\delta_c=1.69$ and $\sigma(M)$ is the rms fluctuation in the matter
density smoothed with a top-hat filter on a scale $R=(3M/4\pi \bar{\rho})^{1/3}$.
We use \cite{SheTor}
\begin{equation}
  b(\nu) = 1 + {\nu-1\over\delta_c} + {2p\over\delta_c(1+\nu'^{p})}
\end{equation}
and
\begin{equation}
  \nu f(\nu) = A(1+\nu'^{-p}) \nu'^{1/2} e^{-\nu'/2}
\label{fnu}
\end{equation}
where $p=0.3$ and $\nu'=0.707\nu$.  The normalization constant $A$ is fixed
by the requirement that all of the mass lie in a given halo
\begin{equation}
  \int f(\nu) d\nu = 1.
\end{equation}

On small scales pairs lying within a single halo become dominant
\begin{equation}
  P^{P}(k) = {1\over (2\pi)^3} \int f(\nu) d\nu
  \ {M(\nu)\over\bar{\rho}} |y(k)|^2.
\label{Poisson}
\end{equation}
The total power spectrum is the sum of the two contributions,
\begin{equation}
P_{\rm dm}(k)=P^{hh}_{\rm dm}(k)+P^P_{\rm dm}(k).
\end{equation}

For galaxies the above model needs to be modified in several 
aspects. Instead of the dark matter particles we are now counting 
galaxies inside halos. Small halos cannot host very bright galaxies, 
so there is a lower mass cutoff in the halo distribution at a given 
luminosity cutoff.
In addition, number of galaxies inside halo need not grow 
linearly as a function of halo mass. Both of these features can be accounted
for by introducing two galaxy multiplicity 
functions, $\langle N \rangle(M)$ and 
$\langle N(N-1)\rangle^{1/2}(M)$, which count mean number of galaxies 
inside the halo both linearly and pair weighted, respectively. 
Second modification is that radial profile of galaxy distribution, $y_{\rm g}(k)$,
need not be the same as that of the dark matter $y(k)$. Both observations 
\cite{Carl} and numerical simulations \cite{Dia} show that at least for
some types of galaxies the two functions must differ. Here we will
for the most part adopt the approach where the two are equal, but 
we also discuss the modifications when this assumption is dropped. Finally, 
one expects there will be one galaxy which forms at the center of the halo. 
The correlations between this 
galaxy and the rest of the galaxies inside the halo
will be sensitive only to a single convolution in the 
radial profile. For large halos with $\langle N(N-1) \rangle^{1/2} \gg 1$ 
the presence of the central galaxy does not change significantly the 
number of pairs or their statistics. 
For small halos where $\langle N(N-1) \rangle^{1/2} \ll 1$
its existence changes the correlations significantly and is in fact 
necessary to explain the steep power law in the galaxy correlation 
function to small scales \cite{Sel,PeaSmi}. 

Putting the above together we have,
\begin{equation}
P^{hh}_{\rm gg}(k)=P_{\rm lin}(k) \left[{\bar{\rho}\over \bar{n}}\int
f(\nu)d\nu  {\langle N \rangle \over M}b(\nu)y_g(k,M)\right]^2,
\label{ghh}
\end{equation}
where $\bar{n}$ is the mean density of galaxies in the sample,
\begin{equation}
 \int {\langle N \rangle \over M} f(\nu)d\nu={\bar{n} \over \bar{\rho}}.
\label{ngal}
\end{equation}
The Poisson term is given by
\begin{equation}
P^P_{\rm dm}(k)= {1 \over (2\pi)^3 \bar{n}^2}
\int {M \over \bar{\rho}}f(\nu)d\nu {\langle N(N-1)\rangle \over M^2}
|y(k,M)|^p,
\label{gp}
\end{equation}
where we approximate the effect of the central galaxy by using 
$p=1$ for $\langle N(N-1)\rangle < 1$ and $p=2$ otherwise.

In redshift space there are two effects which modify the above expressions.
The first is a boost of power on large scales due to streaming
of matter into overdense regions \cite{Kai}.  
The second is a reduction of power on small
scales due to virial motions within an object \cite{PD}.  
In the halo model these
two effects can be separated into the halo-halo and one halo (Poisson) 
contributions. In this model the 
virial motion suppression becomes a function of scale, since larger
halos (with larger velocity dispersions) dominate at larger scales 
than smaller halos. In contrast to the previous models \cite{Hat}
this model succesfully reproduces N-body simulation results \cite{White}.

In linear theory a density perturbation $\delta_k$ generates a velocity
perturbation $\dot{\delta}=-ikv$ with $\vec{v}$ parallel to $\vec{k}$.
Using the plane-parallel approximation ($kr\gg 1$)
the redshift space galaxy density perturbation $\delta^{\rm rs}_{g}$ 
is given by
\begin{equation}
  \delta^{\rm rs}_{\rm g} = \delta_{\rm g}+\delta_{\rm v}\mu^2,
\label{lsrd}
\end{equation}
where $\mu=\hat{r}\cdot\hat{k}$, $\delta_{\rm g}$ is the real space
galaxy density perturbation and $\delta_{\rm v}$ is the 
velocity divergence. This can be related to the density perturbation 
$\delta_{\rm dm}$ via $\delta_{\rm v}=f\delta_{\rm dm}$, 
where $f(\Omega)\equiv d\log\delta/d\log a\simeq \Omega_m^{0.6}$ and $a$ is the
scale-factor. 

On small scales virial motions within collapsed objects act as a 
gaussian convolution in 
redshift space, which suppresses power. We will model this as
a gaussian filter with mass dependent 1-d velocity dispersion $\sigma$, 
acting on the mode component along the line of sight.
Assuming that the halos are isothermal we may use the mass within the virial
radius to obtain the 1D velocity dispersion of a halo of mass $M$, 
\begin{equation}
  \sigma=[GM/2r_{\rm vir}]^{1/2} \sim 7 H_0r_{\rm vir}, 
\end{equation}
where $H_0$ is the Hubble constant and $r_{\rm vir}$ is the virial 
radius, which can be related to the halo virial mass, $M_{\rm vir}=
4 \pi \delta_{\rm vir}\bar{\rho}r_{\rm vir}^3/3$.The density contrast in 
redshift space is
\begin{equation}
  \delta^{\rm rs}_{\rm g} = \delta_{\rm g} e^{-(k\sigma\mu)^2/2}.
\label{dr1dmu}
\end{equation}

Since the power spectrum is in general a function of both $k$ and $\mu$ 
we must decide which quantity we are interested in before proceeding. 
The most common is 
to average over $\mu$ to obtain the isotropized power spectrum. 
On large scales this implies 
averaging over $\mu$ the square of equation \ref{lsrd}.
This can be further improved by including 
the small scale dispersion. A choice that seems to work well in 
comparison to simulations is to add a term obtained by radially averaging 
equation \ref{dr1dmu} (White 2000).
On small scales we only need to 
average over the square of equation \ref{dr1dmu}, 
since there is no linear effect. 

Combining the above 
the isotropized redshift space power spectrum in the halo model 
becomes
\begin{eqnarray}
& &P_0(k) = \left( F_{\rm g}^2+{2\over 3}F_{\rm v}F_{\rm g} + {1\over 5}F_{\rm v}^2\right)P_{\rm lin}(k)
   \\
& & + 
{1 \over (2\pi)^3 \bar{n}^2}
\int {M \over \bar{\rho}}f(\nu)d\nu {\langle N(N-1)\rangle \over M^2}
{\cal R}_p(k\sigma) |y_{\rm g}(k,M)|^p,
\nonumber
\label{rs}
\end{eqnarray}
where 
\begin{eqnarray}
F_{\rm v}&=&f\int f(\nu) d\nu\ b(\nu) {\cal R}_1(k\sigma) y(k; M) 
\nonumber \\
F_{\rm g}&=&
{\bar{\rho}\over \bar{n}}\int
f(\nu)d\nu  {\langle N \rangle(M) \over M}b(\nu){\cal R}_1(k\sigma)y_{\rm g}(k,M)
\label{p0}
\end{eqnarray}
and 
\begin{equation}
  {\cal R}_p(\alpha=k\sigma[p/2]^{1/2}) = {\sqrt{\pi}\over 2} { {\rm erf}(\alpha)\over \alpha},
\label{r2}
\end{equation}
for $p=1,2$.

\begin{figure}
\begin{center}
\leavevmode
\epsfxsize=8cm \epsfbox{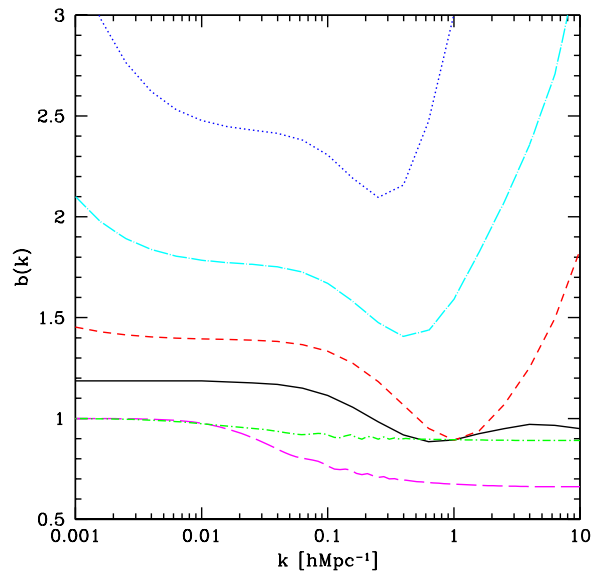}
\end{center}
\caption{Bias $b(k)$ defined as the square root of
the ratio of redshift space nonlinear power spectrum to the 
linear real space dark matter spectrum. From top to bottom are shown 
very red galaxies ($M_B-M_V>2$, dotted), red galaxies ($M_B-M_V>0.8$, 
long dash-sotted), 
normal galaxies (short dashed) and dark matter (solid). The bottom two lines are 
ratios between massive and massless neutrino
transfer functions, for $m_{\nu}=0.1eV$ (upper, short dash-dotted)
and $m_{\nu}=1eV$ (lower, long-dashed).
}
\label{fig1}
\end{figure}

Figure \ref{fig1} shows various bias functions as a function of $k$, 
defined as the square root of the 
ratio of the galaxy to the linear dark matter power spectrum. 
Throughout we use
$\Lambda CDM$ model
with $\Omega_m=0.3$, $\Omega_{\lambda}=0.7$,
normalized to $\sigma_8=0.9$ today.
Linear dark matter power spectrum is used here, since this 
is the quantity we wish to reproduce. 
Also shown is the square root 
ratio of the nonlinear redshift space dark matter spectrum to the linear 
power spectrum. This quantity can be 
compared directly to the N-body simulations. We verified it to be in a
remarkable agreement with those (see also White 2000).
On large scales it should correspond to $b=1.18$.
It is interesting to note that the redshift space dark matter 
power spectrum agrees to within 20\% with the real space linear 
power spectrum over the entire range of scale. This is 
somewhat coincidental, since for the nonlinear redshift space 
power spectrum the correlated halo-halo term is 
suppressed on small scales because of virial motions $({\cal R}_1(k\sigma)<1$)
and because of halo profile ($y(k;M)<1$).
The difference is picked up 
by the Poisson term, which does not depend on the linear power 
spectrum, except through the mass dependence of the 
concentration parameter.  

For galaxies we see that the linear bias, defined as the ratio 
of the redshift space galaxy power spectrum to the linear dark matter 
spectrum, typically 
exceeds the redshift space dark matter bias on large scales. 
This is of course not surprising and reflects the fact that 
galaxies are a biased tracer of dark matter. For galaxies 
found predominantly in groups and clusters the bias is larger
than for those which are also found in the field. 
The choice of the galaxy multiplicity function used here is motivated by 
the semi-analytic models of galaxy formation \cite{Ben,Kau}. Least biased
are regular galaxies selected only on the 
basis of their luminosity (dashed line). 
More biased are red galaxies with $M_B-M_V>0.8$ in semi-analytic 
models of Kauffmann et al. (1999; dash-dotted), while most biased 
are those with $M_B-M_V>2$ (dotted). It is important to emphasize 
that these are just
plausible choices of the galaxy multiplicity function 
as the data at 
present do not allow one to determine these directly. 
Theoretical models of galaxy multiplicity function 
can vary at least at the level of 30\%, leading to 
variations of up to 150 km/s in 1-d velocity dispersion.
For this reason we only emphasize the features of the model 
which are generic and expected for any choice of the parameters, 
even though the relative importance of different effects may vary 
from model to model.
For example, in all cases the galaxy multiplicity function is
expected to increase
less rapidly with the halo mass than the mass itself, but may have a
different low mass cutoff and/or different shape.

The bias is scale independent on large scales for the regular 
galaxies. For these
bias becomes scale dependent above $k>0.1h$Mpc$^{-1}$, on scale
somewhat smaller than for the dark matter. This is because these galaxies
are preferentially found in smaller halos relative to the dark matter, 
and are hence dominated by systems with smaller $\sigma$, thus
the virial motion suppression is smaller. Just as in the case of
the dark matter the galaxy linear bias first declines 
with scale.  
This is because correlations are suppressed both by the finite extent
of the halos and by the virial motions within them. 
At even smaller scales, above $k>1h$Mpc$^{-1}$,
bias begins to rise again. This is caused by the nonlinear 
Poisson term for the galaxies, not present in the linear power spectrum, 
which enhances the correlations
on small scales. This is of course the term that gives rise to the 
nonlinear clustering pattern and the familiar power law slope.
Just as in the real space in the redshift space
the enhancement is also more important for the galaxies
than for the dark matter.
One reason for this is that galaxies are
preferentially found in smaller systems relative to the dark matter, 
so the suppression because of finite halo extent and virial motions 
is less important than for the dark matter. Another is that on 
smaller scales the correlations are dominated by 
small halos with $\langle N \rangle <1$. 
In these there is a central galaxy, which 
does not contribute to the suppression because of halo profile and 
virial motions, leading to an enhancement of the galaxy power spectrum 
over that of the dark matter. Finally, 
a more concentrated 
distribution of galaxies relative to the dark matter, as suggested by
some observations \cite{Carl}, would also lead to an enhancement 
of the galaxy power spectrum relative to that of the dark matter. 

The choices motivated by red or elliptical galaxies 
in semi-analytic models \cite{Kau} 
elliminate galaxies in less massive halos.
These are shown in figure \ref{fig1} as 
the dash-dotted and dotted lines, for the case of red ($M_B-M_V>0.8$) and 
very red ($M_B-M_V>2$) galaxies, respectively. 
The bias in this case shows scale dependence even on very large scales, 
$k<0.01h$Mpc$^{-1}$. This is because the Poisson term becomes 
important again relative to the linear power spectrum on very large 
scales, where the latter approaches slope $ n \sim 1$, while 
the former remains at $n=0$. Galaxies that are only
found in rarer, more massive, systems have 
the Poisson term that is larger, so this effect 
is relatively more important for this type of galaxies as opposed to the
normal galaxies. As pointed out in Seljak (2000) this effect is not 
present for dark matter because of mass and momentum conservation, 
which requires the power spectrum of a local process to decline as 
$P(k) \propto k^4$ on large scales. Since the Poisson term is constant 
on large scales one can attempt to model it as such and remove it. Note 
that its amplitude is larger than the usual Poisson term arising from 
the discretness of galaxies, which depends on the total number of 
galaxies. Here the amplitude of Poisson term is determined by the effective
number of halos, which is smaller than the number of galaxies.
If one removes this constant term then the remaining power spectrum 
should trace dark matter, so bias should become scale independent again.

For the same reason that these galaxies are in more 
massive halos the virial motion and finite halo size suppression 
become important on larger scales and the bias is already scale dependent
for $k \sim 0.1h$Mpc$^{-1}$. On smaller scales the bias dependence is also
sensitive to whether the galaxies are central inside the halo or not and 
whether the radial distribution follows that of dark matter, both 
of which can have important impact on the bias. For example, 
if red galaxies avoid centers of halos (because the gas preferentially 
cools to the halo center, where it can form new stars which are blue), then 
the bias is a less rapidly rising function of $k$ on small scales  than 
that shown in figure 
\ref{fig1}).

\section{{\bf $\beta$} from redshift distortions}

Another application of the redshift distortion analysis is to 
extract the parameter $\beta=f/b$. There are many different ways to do 
this (see Strauss \& Willick 1995 for a review). 
The two examples used here are by combining the 
galaxy power spectrum $P_{\rm gg}$ with the velocity power spectrum 
$P_{\rm vv}$ or their cross-spectrum $P_{\rm vg}$ and by using 
Legendre expansion of redshift space power spectrum. If bias 
is constant on large scales $\beta$ will also be a constant, but 
in the more general case considered here it will be scale dependent. 
Moreover, different methods may not even agree on a given scale, so 
the meaning of $\beta$ itself becomes questionable.

Given that in general the power spectrum is a function of both 
$\mu$ and $k$ it is not possible to extract the three components 
$P_{\rm gg}$, $P_{\rm vg}$ and $P_{\rm vv}$
from it uniquely, since it depends on the adopted procedure. 
In fact, a clean separation into the three components is 
only possible on large scales where only the linear compression redshift
distortion operates. To obtain an idea what happens when small scale effects 
become important we perform angular averages assuming linear 
theory model and reconstruct the 3 components from these. 
As before we assume plane-parallel approximation.
For isotropic averaging in equation \ref{p0} this gives
\begin{equation}
P_0(k)=
{1\over 2}\int_{-1}^{+1}d\mu\ \left(\delta_g+ \mu^2\delta_v\right)^2 =
P_{\rm gg}+{2 \over 3}P_{\rm vg}+{1 \over 5}P_{\rm vv}.
\end{equation}

Similarly we can also perform averaging 
with $\mu^2$ and $\mu^4$
weights, 
$$
P_2(k)=
{1\over 2}\int_{-1}^{+1}\mu^2d\mu\ \left(\delta_g+ \mu^2\delta_v\right)^2 =
{1 \over 3}P_{\rm gg}+{2 \over 5}P_{\rm vg}+{1 \over 7}P_{\rm vv} 
$$
\begin{equation}
P_4(k)=
{1\over 2}\int_{-1}^{+1}\mu^4d\mu\ \left(\delta_g+ \mu^2\delta_v\right)^2 =
{1 \over 5}P_{\rm gg}+{2 \over 7}P_{\rm vg}+{1 \over 9}P_{\rm vv}. 
\end{equation}
From $P_0$, $P_2$ and $P_4$ we can reconstruct uniquely $P_{\rm gg}$, 
$P_{\rm vg}$ and $P_{\rm vv}$ using the above expressions. 
At the same time the halo model also gives predictions for the
averaged spectra $P_2$ and $P_4$ just like for $P_0$ (equation \ref{p0}). 
These are given by
\begin{eqnarray}
& &P_2(k) = \left( {1 \over 3}F_{\rm g}^2+{2\over 5}F_{\rm v}F_{\rm g} + {1\over 7}F_{\rm v}^2\right)P_{\rm lin}(k)
   \nonumber \\
& &+ 
{1 \over (2\pi)^3 \bar{n}^2}
\int {M \over \bar{\rho}}f(\nu)d\nu {\langle N(N-1)\rangle \over M^2}
{\cal R}_{p+2}(k\sigma) |y_{\rm g}(k,M)|^p \nonumber \\
& &P_4(k) = \left( {1 \over 5}F_{\rm g}^2+{2\over 7}F_{\rm v}F_{\rm g} + {1\over 9}F_{\rm v}^2\right)P_{\rm lin}(k)
   \nonumber \\
& & + 
{1 \over (2\pi)^3 \bar{n}^2}
\int {M \over \bar{\rho}}f(\nu)d\nu {\langle N(N-1)\rangle \over M^2}
{\cal R}_{p+4}(k\sigma) |y_{\rm g}(k,M)|^p \nonumber, \\
\label{eqns}
\end{eqnarray}
where 
\begin{equation}
{\cal R}_{p+q}(k\sigma)=\int_{0}^{1} \mu^q e^{-p(k\sigma\mu)^2/2} d\mu,
\end{equation}
which have simple analytic expressions similar to equation
\ref{r2}, 
\begin{equation}
{\cal R}_{p+2}(\alpha=k\sigma[p/2]^{1/2}) = 
{\sqrt{\pi}\over 4}\left[ { {\rm erf}(\alpha)\over \alpha^3}-{e^{-\alpha^2} \over 
2\alpha^2} \right]\nonumber 
\end{equation}
$$
{\cal R}_{p+4}(\alpha=k\sigma[p/2]^{1/2}) = 
{3\sqrt{\pi}\over 8} \left[ { {\rm erf}(\alpha)\over \alpha^5}-{e^{-\alpha^2} \over 
2\alpha^2}\left(1+{3 \over 2 \alpha^2}\right)\right],
$$
for $p=1,2$.

\begin{figure}
\begin{center}
\leavevmode
\epsfxsize=8cm \epsfbox{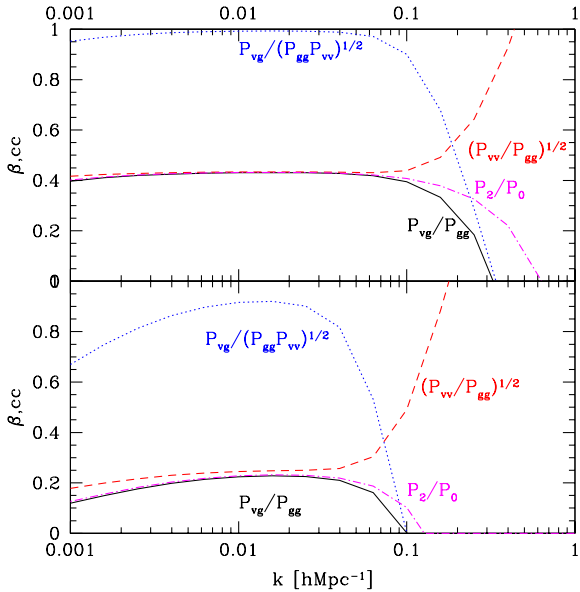}
\end{center}
\caption{The predictions of the halo model for $\beta$ using 
a variety of methods discussed in the text and galaxy-velocity
cross-correlation coefficient. Top figure shows regular galaxies
found also in the field, bottom shows biased (red, elliptical etc.)
galaxies found predominantly in groups and clusters.}
\label{fig2}
\end{figure}

The procedure to extract $\beta$ is the following: first we
compute $P_0$, $P_2$ and $P_4$ from above expressions. Next 
we assume they are determined by the
linear combinations of $P_{\rm gg}$, $P_{\rm vg}$ and $P_{\rm vv}$
as valid in linear theory, which allows one to determine
them uniquely. 
We then take ratios $[P_{\rm vv}/P_{\rm gg}]^{1/2}$ and 
$P_{\rm vg}/P_{\rm gg}$ to determine $\beta$. 
These results 
are shown in figure \ref{fig2} for normal (top) and 
red ($M_B-M_V>0.8$; bottom) galaxy sample used in figure 
\ref{fig1}. We see that for normal galaxies the two 
reconstructed $\beta$ functions
are approximately constant and equal for $k<0.1h$Mpc$^{-1}$. 
For $k>0.1h$Mpc$^{-1}$
the two $\beta$ functions diverge away from the large scale
value and away from 
each other. This indicates that on scales below 50$h^{-1}$Mpc one
cannot extract the true value of $\beta$ and that different 
methods of determining it can give rather different answers.
This is because on these scales virial motions within halos 
become important and pure linear theory ansatz is no longer valid.
The scale dependence of the two $\beta$ reconstructions is in 
qualitatively good agreement with the behaviour seen in real 
data. For example, \cite{Ham} have decomposed the data in 
a similar way to $P_{\rm gg}$, $P_{\rm vg}$ and $P_{\rm vv}$.
Their reconstructed scale dependence of 
$\beta$ show a similar behaviour
as our model, where $\beta$ from $P_{\rm vg}/P_{\rm gg}$ declines 
with $k$, 
while that from $(P_{\rm vv}/P_{\rm gg})^{1/2}$ increases
with k (see their figure 4). We caution that this comparison is just 
illustative, since the two methods of
analysis differ in details (such as the use of 
plane-parallel approximation) and so cannot be directly compared.

We can also extract the cross-correlation coefficient 
$r=P_{\rm vg}/[P_{\rm vv}P_{\rm gg}]^{1/2}$ from this 
analysis. For normal galaxies it is close to unity for
$k<0.1h$Mpc$^{-1}$ and rapidly declines above that. 
This means that the galaxy density and velocity divergence 
as reconstructed from this method
become poorly correlated on small scales. Given this
it is meaningless to 
combine the different estimates of $\beta$ to enhance the 
statistical significance, since they do not measure the 
same parameter and the two fields are only poorly correlated 
on small scales. 

For biased (red or elliptical) 
galaxies found predominantly in
groups or clusters
the correlation is smaller than unity 
even on large scales and the agreement between different 
reconstructed $\beta$ is less good there. This is mostly caused by the 
Poisson term, which is not negligible for galaxies
on large scales. Divergence between $\beta$ from different methods appears
already for $k \sim 0.04h$Mpc$^{-1}$, caused by the more significant
influence of massive halos with larger velocity dipersions.
One should therefore be specially careful when drawing conclusions 
on cosmological parameters from this sample of galaxies, as for 
example from the planned bright red galaxy sample (BRG) in SDSS.

Another often used way to extract $\beta$ is to 
determine the ratio of quadrupole to monopole terms \cite{Kai}, 
which in terms of the above quantities is given by 
$P_{qm}=2.5(3P_2/P_0-1)$. In linear theory one 
predicts it to be \cite{Cole}
\begin{equation}
P_{qm}={ 1+2\beta/3+ \beta^2/5\over 4 \beta/3+4\beta^2/7}.
\end{equation}
By solving the quadratic equation above one can determine
$\beta$ as a function of scale. This quantity is also 
shown in figure \ref{fig2} and has a similar behaviour to 
$\beta$ from $P_{\rm vg}/P_{\rm gg}$ ratio, although the 
suppression on small scales is delayed relative to it.

\section{Conclusions}
We have analyzed the redshift distortion effects using the halo model,
which has proven to be remarkably succesful in explaining 
nonlinear real space power 
spectrum of both galaxies and dark matter \cite{MaFry,PeaSmi,Sco,Sel},
as well as redshift space power spectrum of the dark matter \cite{White}. 
We have shown that redshift space 
bias for regular galaxies is likely to be constant only for 
$k<0.1h$Mpc$^{-1}$. Between $0.1h$Mpc$^{-1}<k<1h$Mpc$^{-1}$ the 
bias declines by 10-20\% and rises again above that. 
This may be important for attempts to extract the value of 
cosmological parameters from such measurements \cite{Tegmark}, 
which typically assume nonlinear effects are negligible 
up to $k \sim 0.2-0.3h$Mpc$^{-1}$. 

One parameter that is particularly sensitive to this effect is
neutrino mass, which 
suppresses the power spectrum roughly in the same scale range where 
bias also becomes scale dependent. Since it has the same effect the 
two can be degenerate over the quasi-linear regime 
and this would complicate the attempts of 
accurately determining the neutrino mass from such galaxy 
clustering measurements. Figure \ref{fig1} shows that the 
effect of neutrino mass begins to affect the power spectrum already 
on scales larger than where the bias is scale dependent. This is 
good news for the efforts to extract the neutrino mass, but the 
effect can only be measured if very large scale correlations can 
be accurately measured. These are limited by finite volume sampling variance,
which can only be reduced by having a larger volume, so that the 
amplitude of correlations can be reliably determined for 
$k \ll 0.1h$Mpc$^{-1}$. If only the information around $k \sim 0.1h$Mpc$^{-1}$
is used then one {\it cannot} separate the scale dependent bias
effects from those of massive neutrinos. Current surveys such as PSCz
still have large errors for $k \ll 0.1h$Mpc$^{-1}$ and the new generation 
of surveys, such as SDSS and 2dF, combined with a more detailed modelling,
will be needed to determine the neutrino mass from these measurements.

Galaxies found in rare systems such as groups and clusters
suffer from another effect. For such galaxies the Poisson term 
can be so strong that it can exceed the linear correlation term
not only on small scales, but also on very large scales. This effect is 
not present for dark matter, which is protected from it by causality
and conservation of mass and momentum \cite{zeldovich}. Because of this 
the bias can rise 
again for $k<0.01h$Mpc$^{-1}$. Since on such large scales this is 
a pure Poisson term with the slope exactly $n=0$ one can attempt to model it
as a sum of two contributions and its amplitude can be estimated from the
small scales. Equivalently, one can perform the correlation function analysis, 
where this term is not present on large scales. 

Determination of parameter $\beta$ shows similarly a converegence
to a single value for $k<0.1 h$Mpc$^{-1}$, at least for normal galaxies
where the Poisson term does not become important on very large scales.
Around and above $k \sim 0.1 h$Mpc$^{-1}$ $\beta$ rapidly becomes 
scale dependent. The actual behaviour depends on the specific analysis 
and $\beta$ can either grow or decline with scale. This is caused by 
the effect of 
virial motions, which counter the linear compression effect and are
more important in more massive halos, which dominate the nonlinear clustering 
on large scales. The model reproduces well the scale dependence of 
$\beta$ seen in the analysis of the real data (Hamilton et al. 2000). 
The cross-correlation coefficient between the galaxy and the velocity field
divergence
as obtained from the redshift space distortions is close to unity 
below $k \sim 0.1 h$Mpc$^{-1}$ and rapidly declines above that.
Galaxies and velocities are singificantly less well correlated than 
the dark matter and the galaxies in real space analysis, where the 
cross-correlation coefficient is unity at least up to 
$k \sim 1 h$Mpc$^{-1}$ \cite{Sel}. Our results are in broad agreement
with other recent analysis of nonlinear bias and its effect on 
redshift space distortions and $\beta$ \cite{Hat,Wein}. 
Together these results
argue for a need of a more refined analysis of redshift distortions 
and redshift space power spectrum if the statistical 
power of existing and upcoming redshift space surveys is to be 
fully exploited. This is needed both 
to reduce the systematic effects and to extend the analysis to smaller
scales. Since the halo approach used here reproduces well the results of 
N-body simulations and semi-analytic models it can
serve as a useful framework within which one can extract the true underlying 
cosmological model from the data. 

\section*{Acknowledgments}

I acknowledge the support of NASA grant NAG5-8084 and Packard and Sloan 
Fellowships. I thank SISSA, Trieste,
for hospitality during my visit and Ravi Sheth for 
useful comments and questions which helped improve the manuscript.

\end{document}